# User Manual for the Complex Conjugate Gradient Methods Library
# `CCGPAK` 2.0

Piotr J. Flatau

*Scripps Institution of Oceanography*
*University of California San Diego*
*La Jolla, CA92093*

*email:* `pflatau@ucsd.edu`

last revised: August 23, 2012

**Abstract**

This manual describes the library of conjugate gradients codes `CCGPAK`, which solves system of complex linear system of equations. The library is written in FORTRAN90 and is highly portable. The codes are general and provide mechanism for matrix times vector multiplication which is separated from the conjugate gradient iterations itself. It is simple to switch between single and double precisions. All codes follow the same naming conventions.

This manual should be cited as:

P. J. Flatau, 2012, "User manual for the complex conjugate gradient methods library. CCGPAK", `http://code.google.com/p/conjugate-gradient-lib/`.

## Contents

# 1 Introduction

CCGPAK is designed to solve single or double precision system of linear equations

**Ax = y**

where **A** is complex matrix, **x** is unknown, and **y** is known right hand side vector. The impetus for this library has been provided by applications in computational electromagnetics, in particular in the Discrete Dipole Approximation code DDSCAT (see Piotr J. Flatau, Improvements in the discrete-dipole approximation method of computing scattering and absorption, Optics Letters, 22, 1205-1207). However, the codes are general and not limited to such applications. All of the codes are written in FORTRAN90. Older FORTRAN implementations were converted using NAGWare Fortran Tools, a software package discontinued in 2008 by NAG, which nevertheless is still a very convenient way to standardize declarations, precision, and polish the code. In many cases I have removed calls to BLAS (Basic Linear Algebra Subprograms) in favor of FORTRAN90 intrinsics such as **dot_product** or **sum**. All codes are for complex matrices. It is arguably trivial task to convert them to real system of linear equations but I do not have plans to do so. Clearly, there are now other implementations and libraries of iterative techniques. Most notably one is advised to examine Templates for the Solution of Linear Systems: Building Blocks for Iterative Methods, Richard Barrett, Michael Berry, Tony F. Chan, James Demmel, June M. Donato6, Jack Dongarra, Victor Eijkhout, Roldan Pozo, Charles Romine, and Henk Van der Vorst which is available on the internet as PDF file. The purpose of this library is to provide yet another package which includes recent algorithms or variants.

The copyright © P. J. Flatau is covered by the GNU General Public License including this user guide, unless stated otherwise in the original implementations.

# 2 Codes

Implemented algorithms are from several sources: published libraries, printed algorithms, or converted from other computer languages. There is a chance that some of these algorithms are repetitive or that their implementations are superficially different.

## *2.1 PIM90*

Set of PIMF90 (**bigcg, bicgstab, cgne, cgnr, cgs, chebyshev, qmr, rbicgstab, rgcr, rgmres, rgmresev**) are based on Rudnei Dias da Cunha and Tim Hopkins, 2003, PIM - The Parallel Iterative Methods package FORTRAN90 implementation. Parts of the package were rewritten and most of supplementary routines were removed in favor of FORTRAN90 intrinsics such as dot_product or sum. The documentation is provided in two publications (Rudnei Dias da Cunha, 1995, The parallel iterative Method (PIM) package for the solution of systems of linear equations on parallel computers, Applied Numerical Mathematics, 19, 33-50 and in PIM User Manual). There were reports of errors in FORTRAN77 version of these codes and QMR and TFQMR algorithms are in this version changed.

## *2.2 Sarkar*

These are based on publication of Sarkar et al (T. K. Sarkar, X. Yang, E. Arvas, ''A limited survey of various conjugate gradient methods for complex matrix equations arising in electromagnetic wave interactions'', Wave motion, 1988, 10, 527-546). Programs are: **srcg, rcg, xcg, xpcg, gmcg, smcg, gacg, sacg** and relevant papers are referenced in Sarkar work. These codes were written in Fortran66. They were rewritten in 1993 by P. J. Flatau, T. Schneider and F. Evans then at Colorado State University, Department of Atmospheric Science. They were again rewritten to FORTRAN90 by Piotr J. Flatau in 2012.

## *2.3 ILUCG*

This code **petr** was implemented by B. T. Draine in early version of the Discrete Dipole Approximation code DDSCAT (http://code.google.com/p/ddscat/) in late 1980 and is distributed with DDSCAT ever since. It was rewritten by P. J. Flatau to FORTRAN77 and subsequently to FORTRAN90. The algorithm is based on note by M. Petravic and G. Kuo-Petravic, 1979, Journal of Computational Physics, 32, 263-269.

## *2.4 BiCOR-CORS*

These routines are based on publication of Carpenteri et al ( B. Carpentieri, Y.-F. Jing and T.-Z. Huang, 2011, The **BiCOR and CORS** Iterative Algorithms for Solving Nonsymmetric Linear Systems, SIAM J. Sci. Comput. 33, 3020-3036).

# 3 Next releases

## *3.1 BiCGstab(l)*

This is the "vanilla" version **of BiCGstab(l)** as described in PhD thesis of D.R.Fokkema, Chapter 3 (also available asPreprint 976, Dept. of Mathematics, Utrecht University, URL http://www.math.uu.nl/publications/). It includes two enhancements to BiCGstab(l) proposed by G.Sleijpen and H.van der Vorst (G.Sleijpen and H.van der Vorst, Maintaining convergence properties of BiCGstab methods in finite precision arithmetic, Numerical Algorithms, 10, 1995, 203-223) and G.Sleijpen and H.van der Vorst (Reliable updated residuals in hybrid BiCG methods, Computing, 56, 1996, 141-163).

## *3.2 CSYM*

The algorithm is published in Angelika Bunse-Gerstner and Ronald Stover, 1999, On a conjugate gradient-type method for solving complex symmetric linera system, Linear Algebra and its Applications, 287, 1999, 105-123. It was also used in dissertation by Sigrid marion Fisher in 2006.

### 3.3 GPBiCG(m,n)

In a paper by Jun Tang, Yongming Shen, Yonghong Zheng, Dahong Qiu, An efficient and flexible computational model for solving the mild slope equation, Coastal Engineering, 2004, 143-154 a new method GPBiCG (m, n), which is a hybrid of BiCGSTAB and GPBiCG methods was proposed. It was subsequently used in computational electromagnetics by Patrick C. Chaumet, and Adel Rahmani, 2009, Efficient iterative solution of the discrete dipole approximation for magnetodielectric scatterers, Optics Letters, 34, 917-919. The version included in CCGPAK is based on rewritten code provided by Patric Chaumet.

### 3.4 COCR

In the paper by Sogabe and Zhang (Tomohiro Sogabe and Shao-Liang Zhang, 2007, A COCR method for solving complex symmetric linear systems, Journal of Computational and Applied Mathematics 199, 297 – 303) a Conjugate *A*-Orthogonal Conjugate Residual (COCR) method was proposed which canbe regarded as an extension of the Conjugate Residual (CR) method. Numerical examples show that COCR often gives smoother convergence behavior than COCG.

## 4 Documentation

There are several original papers, manuals, internet publications, located in the directory "Documentation", which provide basic information. In particular PIMF90 manual provides basic description of the convenctions which are followed in CCGPAK implementations. The codes are not extensively tested although.

## 5 Useage

Compile using provided **make** file. There are module programs: **ddprecision.f90** and **cgmodule.f90** which need to be set. Module **ddprecision.f90** allows setting precision. We use FORTRAN90 feauture to call proper intrinsic depending on precision and variable type. All the communication to **matvec, cmatvec** is set in **module arraymodule**. User can implement his/her own sparse matrix scheme in matvec and cmatve. Module **cgmodule** is used to define structure which is used to communicate tolerance (**epsilon_err**), maximum number of iterations (**maxit**), and some other parameters documented there. **Cgdriver** is a very simple example of how to exercise many or all of the code. The libraries are separated. Currently they are divided into three groups: **PIM, Sarkar, and cglib**. Directory **todo** contains several codes, librariers, or algorithms which are not yet implemented. Users are encouraged to help with implementations and contribute to CCGPAK.

## 6 Acknowledgements

Rudnei Dias da Cunha provided version of the PIM90 code. Their original layout and proposal for iterative techniques is followed here. The original Sarkar codes were also rewritten by Piotr J. Flatau, Tim Scheider and Frank Evans in 1993 and released of the original version of CCGPAK. In that time Tim Schneider implemented CGSQR and CGSAB algorithms based on the H. A. van der Vorst publication. However, the CCGPAK 1.0 codes were not extensively used and we lost track of them. The original CCGPAK codes were found on the internet in 2012 and I rewrote them to FORTRAN90. B. Carpentieri provided CORS implementation. Bruce Draine implemented PETR code based on Petravic and Kuo-Petravic in Fortran66. This code was heavily used in all versions of computational electromagnetics code DDSCAT. Patric Chaumet provided his corrected version of the PIM QMR code as well as implementation of the algorithm published in Jun Tang et al. 2004 paper. The BiCGSTAB2 algorithm is available on M.A.Botchev's internet side and it is written in Fortran90. Tomohiro Sogabe provided his COCR Fortran77 codes and examples with and without preconditioning.